\documentclass[namedreferences]{SolarPhysics}
\usepackage[optionalrh]{spr-sola-addons} 
\usepackage{graphicx}        
\usepackage{color}           
\usepackage{url}             

\newcommand{\farcs}{\hbox{$.\!\!^{\prime\prime}$}}
\newcommand{\arcsec}{\hbox{$^{\prime\prime}$}}

\newcommand{\etal}{{\it et al.}}

\newcommand{\aap}{    {\it Astron. Astrophys.}}

\newcommand{\apj}{    {\it Astrophys. J.}}
\newcommand{\apjl}{   {\it Astrophys. J. Lett.}}

\newcommand{\jgr}{    {\it J. Geophys. Res.}}

\newcommand{\pasj}{   {\it Pub. Astron. Soc. Japan}}

\newcommand{\solphys}{{\it Solar Phys.}}

\begin{document}

\begin{article}

\begin{opening}

\title{Granular-Scale Elementary Flux Emergence Episodes in a Solar Active Region}

\author{S.~\surname{Vargas Dom\'{\i}nguez}$^{1,2}$\sep
L.~\surname{van Driel-Gesztelyi}$^{2,3,4}$\sep
        L.R.~\surname{Bellot Rubio}$^{5}$}
\runningauthor{Vargas Dom\'{\i}nguez et al.}
\runningtitle{}

   \institute{$^{1}$Departamento de F\'{\i}sica, Universidad de Los Andes, A.A. 4976, Bogot\'a, Colombia.
                     email: \url{s.vargas54@uniandes.edu.co}\\ 
                      $^{2}$ Mullard Space Science Laboratory, University College London, Holmbury St. Mary, Dorking, RH5 6NT, UK\\
                       email: \url{lvdt@mssl.ucl.ac.uk}\\
                     $^{3}$ Observatoire de Paris, LESIA, FRE2461(CNRS), F-92195 Meudon Principal Cedex, France.\\
                     $^{4}$ Konkoly Observatory of Hungarian Academy of Sciences, Budapest, Hungary.\\
              $^{5}$ Instituto de Astrof\'{\i}sica de Andaluc\'{\i}a (CSIC), Apdo. 3004, 18080, Granada, Spain.
                     email: \url{l.bellot@iaa.es} \\
             }

\begin{abstract}
We analyze data from \emph{Hinode} spacecraft taken over two 54-minute periods during the emergence of AR 11024. 
We focus on small-scale portions within the observed solar active region and discover the appearance of very distinctive small-scale and  short-lived dark features in Ca {\sc II}~H chromospheric filtergrams and Stokes $I$ images. The features appear in regions with close-to-zero longitudinal magnetic field, and are observed to increase in length before they eventually disappear. Energy release in the low chromospheric line is detected 
while the dark features are fading. Three complete series of these events are detected having remarkably similar properties, {\it i.e.} lifetime of $\approx$12 min, maximum length and area of 2-4 Mm and 1.6-4 Mm$^{2}$, respectively,
and all had associated brightenings. In time series of magnetograms a diverging bipolar configuration is observed accompanying the appearance of the dark features and the brightenings. The observed phenomena are explained as evidencing elementary flux emergence in the solar atmosphere, {\it i.e}  small-scale arch filament systems rising up from the photosphere to the lower chromosphere with a length scale of a few solar granules. Brightenings are explained as being the signatures of chromospheric heating 
triggered by reconnection of the rising loops (once they reached chromospheric heights) with pre-existing magnetic fields as well as to reconnection/cancellation events in U-loop segments of emerging serpentine fields. The characteristic length scale, area and lifetime of these elementary flux emergence events agree well with those of the serpentine field observed in emerging active regions. We study the temporal evolution and dynamics of the events and compare them with the emergence of magnetic loops detected in quiet sun regions and serpentine flux emergence signatures in active regions. Physical processes of the emergence of granular-scale magnetic loops seem to be the same in quiet sun and active regions being the difference the reduced chromospheric emission in quiet sun attributed to the fact that loops are emerging in a region of lower 
ambient magnetic field density and therefore interactions and reconnection are less likely to occur. Incorporating the novel features of granular-scale flux emergence presented in this study we advance the scenario for serpentine flux emergence. 
\end{abstract}
\end{opening}

\section{Introduction}
     \label{S-Introduction} 

Solar activity involves a complex interplay of processes on many spatial and time scales, fuelled by its internal magnetism. In the current paradigm, the magnetic field generated in the solar interior is brought up to the photosphere where it forms active regions. Emerging flux regions (EFRs) therefore correspond to sites where the magnetic field is breaking through the solar surface exhibiting rising loops \cite{bruzek1969}. The characteristic configuration formed by these magnetic loops is commonly referred to as an arch filament system 
(AFS; \opencite{bruzek1967}). During the last decades many observational studies focussed on flux emergence and its signatures in different solar atmospheric layers. Bright and hot SXR coronal loops are seen above AFSs and EFRs in active regions \cite{kawai1992}. Smaller-scale emerging ephemeral regions show up in SXR observations as X-ray bright points  (\opencite{golub1974}, \citeyear{golub1977}). Flux emergence with increasing shear was proposed to lead to energy release and heating of the overlying corona \cite{deng2000}.

Understanding the emergence of magnetic field from the solar interior is a key topic in solar physics where a combination of ever-increasing temporal and spatial resolution observations and theoretical advances led to significant progress. Different aspects of the response of the solar atmosphere to the emergence of magnetic flux that have been investigated include {\it  e.g.} interaction of flux emergence with the convective plasma, the formation of undulatory or serpentine fields  \cite{strous1999}, undulatory flux emergence associated chromospheric heating \cite{georgoulis2002}, the emergence of serpentine magnetic field and coronal response \cite{harra2010}, emergence into pre-existing ambient field \cite{zuccarello2008,guglielmino2010},  and evolution of photospheric magnetic field and its related coronal response \cite{kubo2003}. The complexity of EFR is also evaluated in detailed numerical simulations studying the evolution of the flux arising from beneath the visible surface  \cite{archontis2008,fan2001}. Small scale mixed-polarity fragments are results of the interaction of emerging flux tubes with the dynamic plasma, as modelled by \inlinecite{cheung2008}, and it has been shown that the situation varies depending on the degree of twist in the rising flux tube, {\it i.e.} weakly twisted flux tubes become more fragmented  (\opencite{magara2004}, \citeyear{magara2006}).
 
Observational signatures of energy release at different scales and heights have been identified above EFRs.  For instance, small-scale brightenings and transient emissions in the wings of the chromospheric H$\alpha$ line, known as Ellerman Bombs or EBs \cite{ellerman1917}, have been detected mainly in active regions. Properties of EBs were studied by {\it e.g.} \inlinecite{socas2006}. In particular \inlinecite{pariat2004}, \inlinecite{pariat2009} proposed that EBs are due to the emergence of resistive serpentine magnetic fields. Theoretical simulations have also been applied to study the formation of EBs associated to the emergence of magnetic flux \cite{archontis2008}. EBs are generally explained as due to reconnection processes changing the morphology of the rising magnetic field lines. Braiding of magnetic field as a result of reconnection has been proposed as a possible mechanism for heating upper layers in the solar atmosphere. Explosive events seems therefore to be triggered by forced magnetic reconnection \cite{jess2010}.
 
Since its launch in September 2006, the {\it Hinode} spacecraft has been widely used to analyze in details the evolution of  EFR. Detailed inspection of the emerging magnetic field and their interactions (with {\it e.g.} ambient fields) as they rise from underneath the visible surface up to the chromosphere, are important to understand key processes of energy release and associated phenomena.  Even in quiet-sun regions, polarimetric observations have evidenced the ubiquitous emergence of magnetic loops \cite{centeno2007,mmartinez2009,gomory2010} and revealed the configuration and dynamics of very small magnetic flux emergence episodes.

In this paper we focus on the study of small-scale phenomena in an emerging solar active region. We detected the emergence of small-scale magnetic loops that appear to interact with pre-existing large-scale magnetic fields. Report on the observations and the data processing are the subjects of Section~\ref{S:observations}. A descriptive overview of the studied solar region is presented in Section~\ref{S:overview}. In Section~\ref{S:description} we focus on the description of the dark features detected in Ca {\sc II}~H in our three studied events. General properties of these emergence episodes is presented in Section~\ref{S:properties}. The discussion and final remarks are presented in Section~\ref{S:discussion}.

\section{Observations and data preparation}
\label{S:observations}

\begin{figure*}
\centering
\includegraphics[angle=0,width=.73\linewidth]{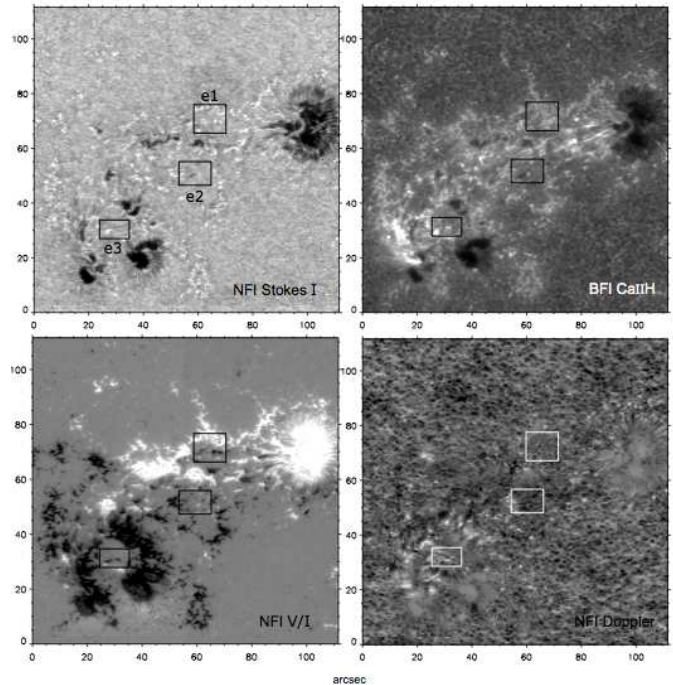} 
\caption{Simultaneous context images of the active region NOAA 11024 observed by \emph{Hinode}/SOT on 4 July 2009  at 18:40 UT.  Contrasted images of Stokes $I$, Ca {\sc ii} H line, NFI Stokes $V$ (with black/white as negative/positive line-of-sight magnetic polarities) and Doppler velocities (where bright features correspond to motions towards the observer) are shown as labeled. Black boxes in all images denote the regions of interest (ROI) analyzed in the present work as correspondingly labeled e1, e2, and e3 in the Stokes $I$ image. Images are contrasted to enhance small-scale faint features.}
\label{context}
\end{figure*}

The Solar Optical Telescope  \cite{tsuneta2008}  onboard the {\it Hinode} satellite \cite{kosugi2007} observed the region NOAA 11024 on 4 July 2009 with the Broadband Filter Imager (BFI) and Narrowband Filter Imager (NFI).  The telescope pointed at solar coordinates ($X$=-55$\arcsec$, $Y$=-488$\arcsec$) at 18:40 UT, where the active region NOAA 11024  (hereafter AR) was located. Filtergrams in the chromospheric spectral line Ca {\sc ii}~H ($\lambda 396.85$ nm) were acquired by the BFI with  pixel size of 0$\farcs$109 and a cadence of 2 min.  The observed field of view (FOV) corresponds to $\approx$112$\arcsec$ $\times$ 112$\arcsec$. The NFI was employed to obtain Stokes $V$ images in the Na photospheric line ($\lambda 5896$ nm), tunable range of 0.6 nm, i.e. $\pm$ 30 pm, Doppler velocity maps and Stokes $I$ images with pixel size of 0$\farcs$16 and a cadence of 2 min\footnote{In this paper we will refer to NFI Na {\sc i}~D Stokes data as NFI Stokes $I$ and NFI Stokes $V$/$I$, not to be confused with the BFI Ca {\sc ii}~H filtergrams.}.  

The observations used for this work were taken from 18:40 to 21:13 UT with a gap of  $\approx$40 min starting at 19:34 UT. The SOT images were corrected for dark current, flat field and cosmic rays by using the standard IDL \emph{SolarSoft} routines. Interpolation was used to replace a few individual corrupted images  ({\it e.g.} due to data packet losses). We end up with two time series of 28 images (54 min) each. Furthermore, a subsonic filter was applied over the Stokes $I$ and Ca {\sc ii}~H sequences to get rid of high frequency oscillations (e.g $p$-modes) in the Fourier space \cite{title1986}.  NFI images were re-sampled to the pixel size of the the BFI data and all the sequences were finally aligned to sub-pixel level and trimmed.

Figure~\ref{context} shows the observations of the AR with black boxes indicating the location of the 
particular events that will be described in detail in this work.

\section{Description of the Analyzed Solar Region}
\label{S:overview}

Active region 11024  shows evidence of twisted flux rope emergence displaying a typical pattern of opposite-polarity magnetic concentrations dubbed "magnetic tongues" {\cite{luoni2011} besides signs of undulatory field (also known as serpentine fields), as comprehensively studied by \inlinecite{valori2012}. The central portion between the two main opposite polarities of the EFR is populated with a wide variety of magnetic elements as seen in the image of circular polarization (NFI V/I) in Figure~\ref{context}. These elements include the well-known MMFs or moving magnetic features \cite{sheeley1969,harvey1973,lee1992} that move at a mean speed of 0.3 - 0.5 km s$^{-1}$ and are found to be a prolongation of penumbral filaments \cite{sainz2005}. MMFs are detached and move away from the two main polarities (sunspot at the top right and dark cores at the lower left corners in the FOV), and are in fact signatures of the decay process, which has already started while the AR is still emerging. Apart from MMFs, a large number of mixed-polarity elements are present along the axis connecting both major positive and negative polarities. Moving bipolar features (MDFs; \opencite{bernasconi2002}) are detected in the central part of the EFR and move at velocities similar to that of the MMFs, though they are likely to drift towards sunspots and not away from them. This intricate pattern of magnetic features is commonly observed in the central part of EFR at the photospheric level \cite{xu2010} and it also is present in output of realistic numerical simulations \cite{cheung2008}.

We aim to investigate small-scale details of magnetic flux emergence in active region AR 11024, which was 
half-way through to its peak evolution and already contained well-developed spots with penumbra. We concentrate on some regions of interest (hereafter ROI) in which small-scale round-shaped dark patches are observed in the Ca {\sc ii}~H images. The events were identified in the \emph{Hinode}/SOT Ca {\sc ii}~H time series and can be described as dark oval areas that appear in the filtergrams, increase their size  and suddenly disappear once their maximum area coverage is reached. 
The fading and eventual disappearance of dark features is accompanied by chromospheric brightenings flanking the dark areas. Immediately after the dark features disappear, an intensification of the bright features is detected around the site where the dark features were previously located. 

Three of these events were markedly detected by visual inspection of the $2 \times 54$-minute time series in regions surrounded (either completely or partially) by more stable, bigger and organized  magnetic areas, as seen in simultaneous magnetograms. The events share comparable overall properties in terms of duration, spatial extent and that they all have associated brightenings. Some other events identified in the time series either did not have full temporal coverage or had less distinctive properties thought they experience a common phenomenology.

\section{Appearance of Dark Features in the Ca {\sc ii}~H Chromospheric Line}
\label{S:description}

The AR studied in this work is populated with a wide variety of solar features including a sunspot with well-developed penumbra; umbrae with partial penumbra; isolated dark cores; abnormal granulation and bright strong field elements, among others (see the Stokes $I$ image in Figure~\ref{context}). The large-scale structure of the region displays a bipolar configuration in the NFI Stokes $V$ magnetogram though individual positive (white) and negative (black) polarities are also mixed over a wide area and it is half-way through its flux emergence process \cite{valori2012}. These mixed-polarity fields have faint circular polarization signals thus indicating weak vertical magnetic fields. The time series evidence a complex interplay of photospheric motions with the embedded magnetic field, leading to the formation of pores and penumbrae. Streaming motions (divergence of opposite polarities) are widely observed in association with global flux emergence. The Ca {\sc ii}~H chromospheric line reveals prominent dynamic with a large number of brightenings and hence intense activity at different spatial scales. 

The ROI analyzed in detail in this work are located in three of the above-mentioned patches. Black boxes e1, e2, and e3 in Figure~\ref{context} denote the location of these ROI. Events e1 and e2 are quite far from major sunspots or stable magnetic entities, whereas e3 is surrounded by some dark pores with developing penumbrae. Nevertheless, in all three cases the NFI Stokes $V$ images reveal intense magnetic activity all around the weak circular-polarization patches where the events are located.
Event e1 takes place in an area of dominantly positive pre-existing field, e3 in negative-field environment, while e2 over the magnetic inversion line of the emerging AR. 

We discovered the appearance and subsequent rapid expansion of dark areas by visual inspection on the Ca {\sc ii}~H time series. The dark features also clearly show up in the Stokes $I$ images a few minutes prior to the Ca {\sc ii} features. After reaching a maximum size, the dark features abruptly disappear. We propose that these dark features correspond to cool absorbing features with dense material, and are manifestations of the flux tubes breaking through the photosphere and their} emergence into the lower chromosphere as small-scale AFSs.
These events provide evidence of the activity and highly dynamic processes at tiny ($<$~10$\arcsec$) spatial scales associated with the overall flux emergence, and are presumably forming part of the large-scale serpentine flux emergence process. In the following subsections we describe the three detected events and analyze their main observational characteristics.
 
\subsection{Event 1}
\label{S:event1}

\begin{figure*}
\centering
\includegraphics[angle=0,width=1.\linewidth]{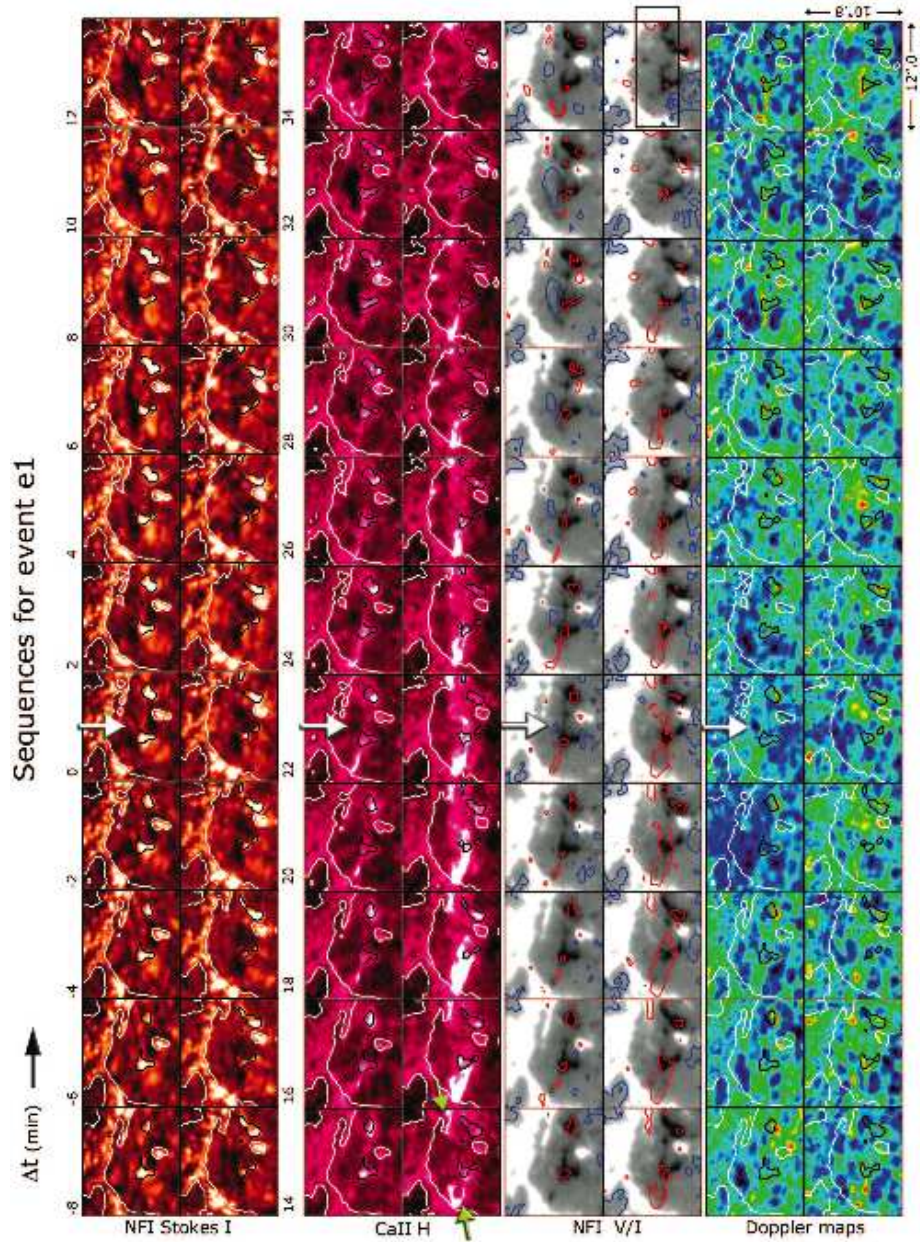} 
\caption{Sequences of contrast-enhanced images for the first event (e1) from \emph{Hinode} (BFI and NFI). From top to bottom: Stokes $I$, Ca {\sc ii}~H, circular polarization V/I  and Doppler images (with blue/red as corresponding to downflows/upflows). The white arrow indicates the onset location of the emerging dark feature. Overlaid black/white contours indicate strong negative/positive circular polarization signals. Blue/red contours in the Stokes $V$ sequence represent dark/bright intensity regions in the Ca {\sc ii}~H sequence. Green arrows point to localized Ca {\sc ii} ~H brightenings. See the text for details. Movie available in the electronic version.}
\label{secuencia1}
\end{figure*}

This event is the largest in extent and most prominent example detected in our data. Figure~\ref{secuencia1} displays the corresponding sequences of simultaneous and co-spatial NFI Stokes $I$, BFI Ca {\sc ii}~H, NFI V/I and Doppler images covering 42 min of observation (from 20:25 UT to 21:07 UT) with 2-min cadence\footnote{See this sequence in movie format as an electronic supplement to this paper.}. The FOV in each panel is 12$\farcs$0$\times$10$\farcs$8. Note that the color convention for Doppler velocities differs from that normally used, so that in our analysis red colour represents upward motions (towards the observer) whereas dark blue colours are downward motions (towards the solar interior). False-colour images are contrasted to enhance very small features which are our main concern for the analysis. Black and white contours (overlaid on Stokes $I$, Ca {\sc ii}~H and Doppler sequences) represent the location of strong negative and positive magnetic polarities extracted from the circular polarization sequence (NFI V/I). Blue and red contours (overlaid on V/I sequence) indicate sites displaying  dark and bright patches observed in the Ca {\sc ii}~H sequence.  

The FOV in the images in Figure~\ref{secuencia1} is dominated by some stable magnetic features that remain practically unchanged during the total observing period. Remarkable unvarying features are for instance the wide positive patches of circular polarization (in white) covering about one third of the FOV and displaying more intense signal in the Stokes $I$ sequence. Another structure, that does not change location but weakens over the 42-min interval, is the much smaller negative element (in black) close to the centre of the FOV. On the other hand, dynamic features are also distinguished.
%


Further inspection of the region allows us to detect a small dark circular patch that starts to be visible in the center of the FOV in the NFI Stokes $I$ and Ca {\sc ii}~H sequences at 08:31 UT and 08:33 UT, respectively, ($\Delta t$=0)\footnote{Initial times $\Delta t$=0 are set as corresponding to the first appearance of the dark feature in all the analyzed sequences.} as indicated by the white arrow in Figure~\ref{secuencia1}. Though the location of appearance of the event displays nearly zero or weak circular polarization signal, it is closely surrounded by positive and negative polarity magnetic fragments, i.e. the stable magnetic features 
mentioned above). In the photosphere (Stokes $I$) there seem to be two parallel narrow dark lanes present under the Ca {\sc ii}~H feature. Though photospheric Doppler maps are very noisy and contaminated by p-modes, prior to the appearance (at  $\Delta t$=-2 ) of the Ca {\sc ii}~H dark features there is a noticeable trend of photospheric downflows in the FOV spread over a larger area in the center/upper part of the FOV. By the time the dark feature appears, the downward photospheric motions are restricted to a well-defined area right at the site of the dark feature. 
In the subsequent images spaced in 2-min intervals, as labeled in the upper sequence in 
Figure~\ref{secuencia1}, the dark feature (outlined with blue contours) turns into an oval structure in Ca {\sc ii}~H  that gets darker and expands up to a maximum size about 10 min after its appearance. A noticeably larger concentration of small-scale brightenings (red contours) is associated with the emergence of the dark feature but they occur in the periphery, i.e., they are not co-spatial. The dark feature eventually fades and disappears. The total duration of the event is of the order of 12 min and, immediately after the dark feature vanishes, intense Ca {\sc ii}~H brightenings are observed on 
both sides of the place where the feature was previously located (see green arrows in Figure~\ref{secuencia1}). The flare-ribbon-like first brightening seen at minute +14 connects the ends of the two dark lanes seen in Stokes $I$. The intensity peak occurs 6 min after the disappearance of the dark feature and in the following images the bright areas gradually shrink and disappear by the end of our observation. 


\subsection{Event 2}
\label{S:event2}

\begin{figure*} 
\centering
\includegraphics[angle=0,width=1.\linewidth]{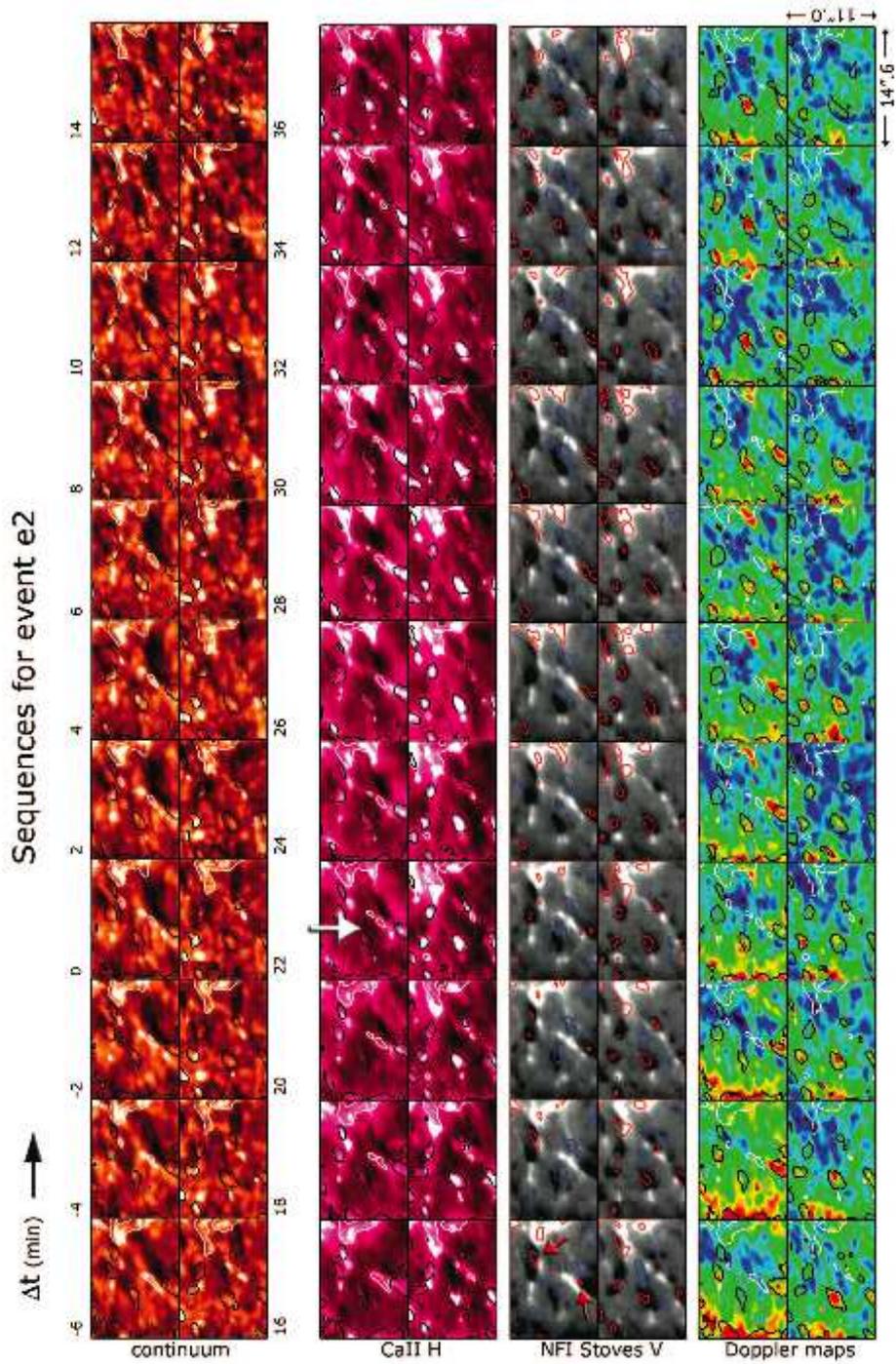} 
\caption{Same as Figure~\ref{secuencia1} for event e2.}
\label{secuencia2}
\end{figure*}

The second example was observed in the region shown in Figure~\ref{context} (e2). Figure~\ref{secuencia2} displays the corresponding sequences of images (following the same analysis as explained in Section~\ref{S:event1} for Figure~\ref{secuencia1}) with individual FOV of 12$\farcs$0$\times$8$\farcs$8. The total duration of the time series is 42 min (from 20:19 to 21:01 UT)\footnote{See this sequence in movie format as an electronic supplement to this paper.}. This is the most complex event in our sample as it displays a large amount of small-scale magnetic features including some MDFs as well as isolated unipolar magnetic polarities. Magnetic concentrations of both polarities are distributed over the FOV, as this event is located over the magnetic inversion line of the AR. 
Same as in the previous event in Section~\ref{S:event1} though slightly smaller, we start to observe a dark patch in the center of the FOV in  Ca {\sc ii}~H sequences in Figure~\ref{secuencia2} at 08:25 UT ($\Delta t$=0) as pointed at by the white arrow in Figure~\ref{secuencia2}. In Stokes $I$  a similar, though somewhat narrower, dark feature appears at the same time.

Remarkably, there is another dark feature observed through the Ca {\sc ii}~H time sequence in Figure~\ref{secuencia2} (about 2$\arcsec$ in the South-West direction from the location of the above-mentioned feature at $\Delta t$=0), that is already visible in the first frame of the series at $\Delta t$=-6 min The two dark features are clearly visible at $\Delta t$=10 min. The dark feature emerging at $\Delta t$=0 (that we will deal with in the following unless stated) seems not to be influenced by the other one, evolving independently with no signs of mutual interaction or dependence. We see evidence of emergence, widening and disappearance. The dark feature gets darker after emerging and expands up to a maximum size of $\approx$~6$\arcsec$~-~7$\arcsec$ in about 10 min after its appearance. A noticeably larger concentration of small-scale brightenings accompanies the disappearance of the dark features (see green arrows in Figure~\ref{secuencia1}). These emerging bipoles show evidence of being part of serpentine fields as indicated by the presence of opposite polarity concentrations adjacent to both polarities. Such configuration is suggestive as a small $\Omega$-loop with two U-loops at each end.   Photospheric upflow features (red-yellow) coincide with the eastern U-loop locations, also marked by bright features, which are not flare-ribbon-like as in e1, but more reminiscent of Ellerman Bombs  (see Figure~\ref{secuencia2}). Global properties of this event and the comparison with event e1 will be deferred to Section~\ref{S:properties}.

\subsection{Event 3}
\label{S:event3}

\begin{figure*}
\centering
\includegraphics[width=1.\linewidth]{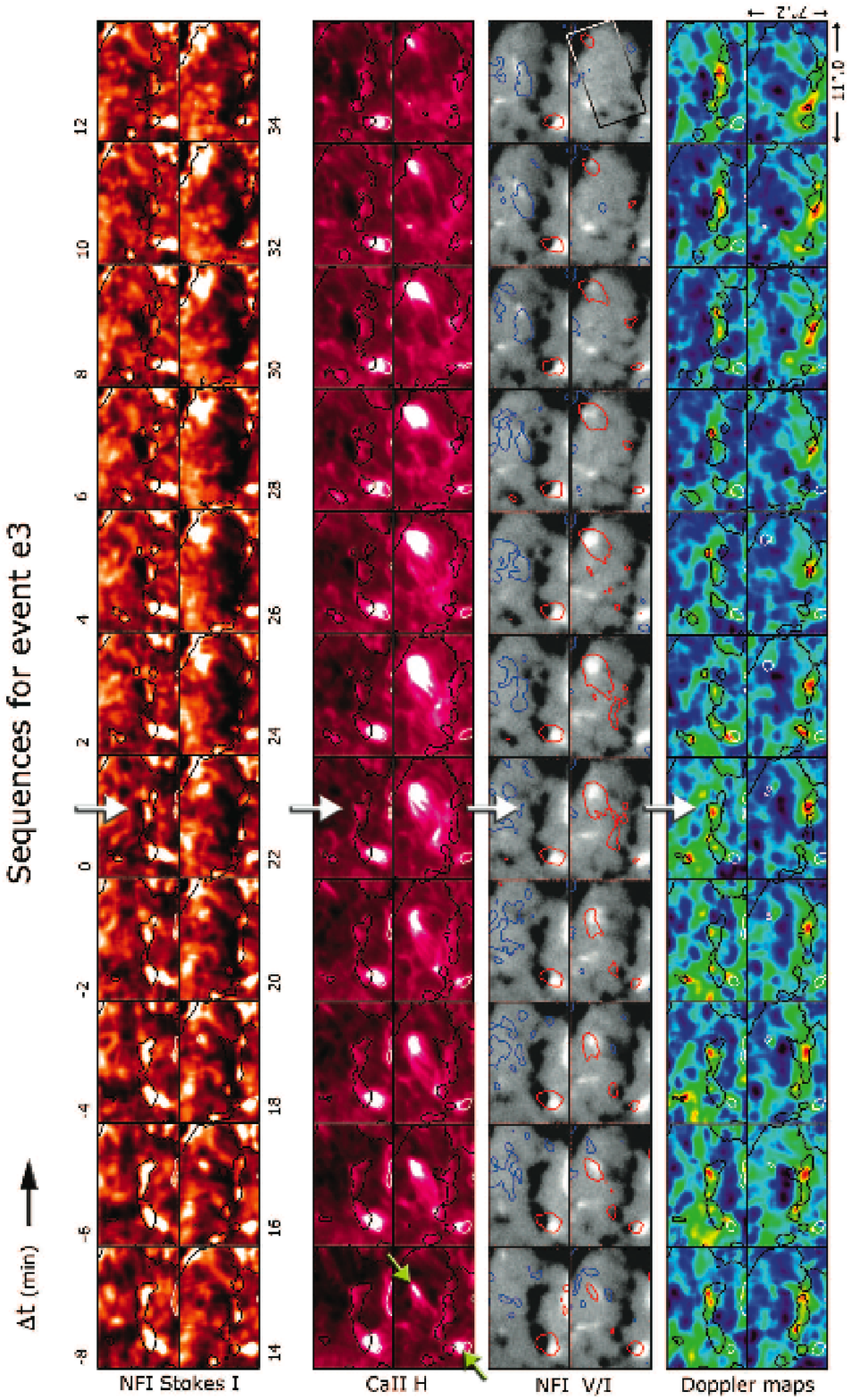} 
\caption{Same as Figure~\ref{secuencia2} for event e3.}
\label{secuencia3}
\end{figure*}

The last case is observed in the region e3 of Figure~\ref{context}, which takes place in the dominantly negative polarity area of the emerging AR. The duration of the sequence shown in Figure~\ref{secuencia3} is 42 min as in the previous two cases (from 18:50 to 19:32 UT) and the FOV in each panel corresponds to 11$\farcs$0$\times$2$\farcs$2. The region is characterized by a predominant negative circular polarization patch placed horizontally across the FOV\footnote{See this sequence in movie format as an electronic supplement to this paper.}. This patch evolves and several individual components separate from each other through the V/I sequence with a detected global drift toward the bottom part of the FOV.

Different from the two previous cases, in this one the onset of the dark feature is a bit diffuse as the initial Ca {\sc ii}~H images are covered by large dark areas ({\it e.g.} see the blue contours in the first four panels in the V/I sequence in Figure~\ref{secuencia3}.) Nevertheless, we identify the emerging dark feature at the location marked by the white arrow in Figure~\ref{secuencia3} ($\Delta t$=0). In the following frames the initially diffuse feature turns into a very well-defined structure that experiences a rapid expansion. This dark feature is actually the most prominent of the ones described in the sections above. The dark feature expands adjacent to a negative circular polarization concentration and lasts for about 12 min before disappearing. A localized downflow is detected at the location of the event which intensifies and remains rather stable during the rest of the sequence.  After the dark feature vanished at $\Delta t$=14 min, a small brightening is detected at one of the ends of the former feature (pointed by the green arrow in Figure~\ref{secuencia3}). A second small brightening appears at $\Delta t$=16 min as pointed by the corresponding green arrow in the figure. Both bright features intensify and develop filamentary structures in between them, {\it i.e.} small loops extending across the area where the dark feature was previously located. The lifetime of the bright structure is about 22 min. In this 3rd case, however, we find no photospheric upflow features co-spatial with the bright ones. There is another close bipole in the SE corner of these images, which is presumably accompanied by an Ellerman Bomb fading by +22 min. In Section~\ref{S:properties} we shall further comment on the properties of the intensity variations.

\subsection{Intensity and area variations of  the dark features}
\label{S:event3}

\begin{figure*}
\centering
\includegraphics[angle=0,width=.93\linewidth]{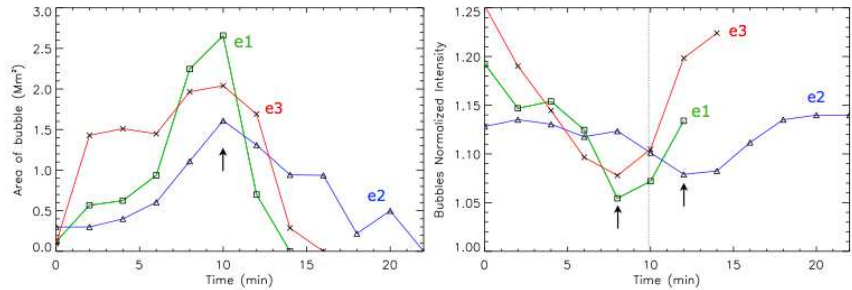} 
\caption{Properties of the dark features for events e1, e2, and e3.  The plots show the variation of area (left panel) and Ca {\sc ii}~H intensity (darkness) (right panel). Values are normalized to a quiet sun area in the intensity profiles. Arrows point 
at the location of maximum area and darkening in the corresponding plots. The vertical line in the plot on the right 
indicates the same time as the arrow on the left plot, for reference.}
\label{intarea}
\end{figure*}

Figure~\ref{intarea} shows the area (left panel) and Ca {\sc ii}~H intensity variation (right panel) with time in the observed dark features for events e1 (squares connected by green line), e2 (triangles connected by blue line), and e3 (crosses connected by red line). The initial time ($\Delta t$=0) is set to the moment where the dark features are first visible in our Ca {\sc ii}~H sequence. The area of the dark features is defined by the number of pixels below the imposed intensity threshold. In the intensity profile the values are normalized to a quiet sun area of  20$\arcsec$$\times$20$\arcsec$ in the lower right part of the FOV in Figure~\ref{context}

The plots evidence a common trend in the time evolution of both quantities. The area starts to increase soon after the first signs of darkness are observed. The area reaches a maximum value 10 min after the appearance of the dark feature (indicated by the arrow on the left panel), which is a common characteristic for all the events. After reaching the peak value, the area decreases until the dark feature eventually vanishes. The declining period is about half  as long as the increasing phase, except for e2 in which both intervals are nearly equally long. Concerning the Ca {\sc ii}~H intensity variations  we found similar results. The initial intensity value ranges from 13 to 25$\%$ higher than the corresponding value over quiet sun. Accompanying their increase in size, the dark features get darker, as their intensity decreases nearly steadily. The minimum intensity values (darkest period) are reached 8 min (e1 and e3) and 12 min (e2) after the appearance of the dark features,  by the time the maximum area is detected  (vertical dashed line on the right plot in Figure~\ref{intarea}). Once the absolute minimum is reached, the profiles show continuous intensity increments. The intensity decrease and increase rates are nearly the same.

These pieces of evidence are suggestive of a cool and dense absorbing feature that crosses the photosphere and is lifted up into a warmer chromospheric medium and hence observed as an expanding dark (absorbing) patch.  The surface area covered by the cool feature will reach a maximum size before its intensity starts to increase suggesting that it is being directly heated. Magnetic loops continue rising while the detected chromospheric dark feature vanishes due to heating and drainage. Such an emerging structure should be detected in data from lower layers where it has already passed through, and that is our main goal for the next section.

\section{Characterizing the Emergence of Small-Scale Arch Filament Systems}
\label{S:properties}

The analysis of chromospheric data has proven to be of key importance to identify the 3-D evolution of the magnetic field that emerges from below. AFS occurring in between the emerging opposite polarities (e.g sunspots, that represent the photospheric footpoints of emerging magnetic loops) are for instance detected in the chromosphere as the ascending dense and cool material contained in the magnetic loops 
\cite{bruzek1967}. In the H$\alpha$ line center dark AFS are observed as dark filaments connecting the magnetic footpoints. Observational and theoretical studies have shed light on many scales of flux emergence happening in the solar atmosphere in both quiet sun and active region. Some very recent studies have revealed the importance of small-scale flux emergence \cite{mmartinez2009,guglielmino2010,otsuji2010} and their consequences in upper atmospheric layers.

The observations on 4 July 2009 display a wide range of interesting phenomena associated to the activity of newly emerging magnetic flux as 
described below and hence our interest on fine scales that allow us to detect magnetic loops or small arch filament systems (SAFS). The appearance of dark features in the Ca {\sc ii}~H sequences was the first sign that drew our attention. These SAFSs are preceded and accompanied by dark photospheric features of about the same length, but somewhat smaller width.
In Section~\ref{S:description} we described the overall configuration of the regions analyzed in this work and commented on the general aspects of the emergence of these dark features. In all detected cases the observed phenomena is essentially the same as far as the morphology  and evolution of the dark features are concerned, i.e. they become visible, increase in size and disappear accompanied by the intensification of chromospheric line emission.

\subsection{Divergence of the Magnetic Bipoles}

From the analysis of the photospheric circular polarization signals recorded by {\it Hinode}/NFI we detected the emergence of magnetic bipoles in the vicinity of the dark features in all of the studied cases.  In this section we deal with the detailed identification of the emerging bipoles, i.e. corresponding positive and negative magnetic features that would indicate the footpoints of emerging loops. Footpoints should be accordingly located on both sides of the dark features found in Section~\ref{S:description} if they truly correspond to an emerging arch filament system at this small spatial scale.

\begin{figure*} 
\centering
\includegraphics[angle=0,width=0.98\linewidth]{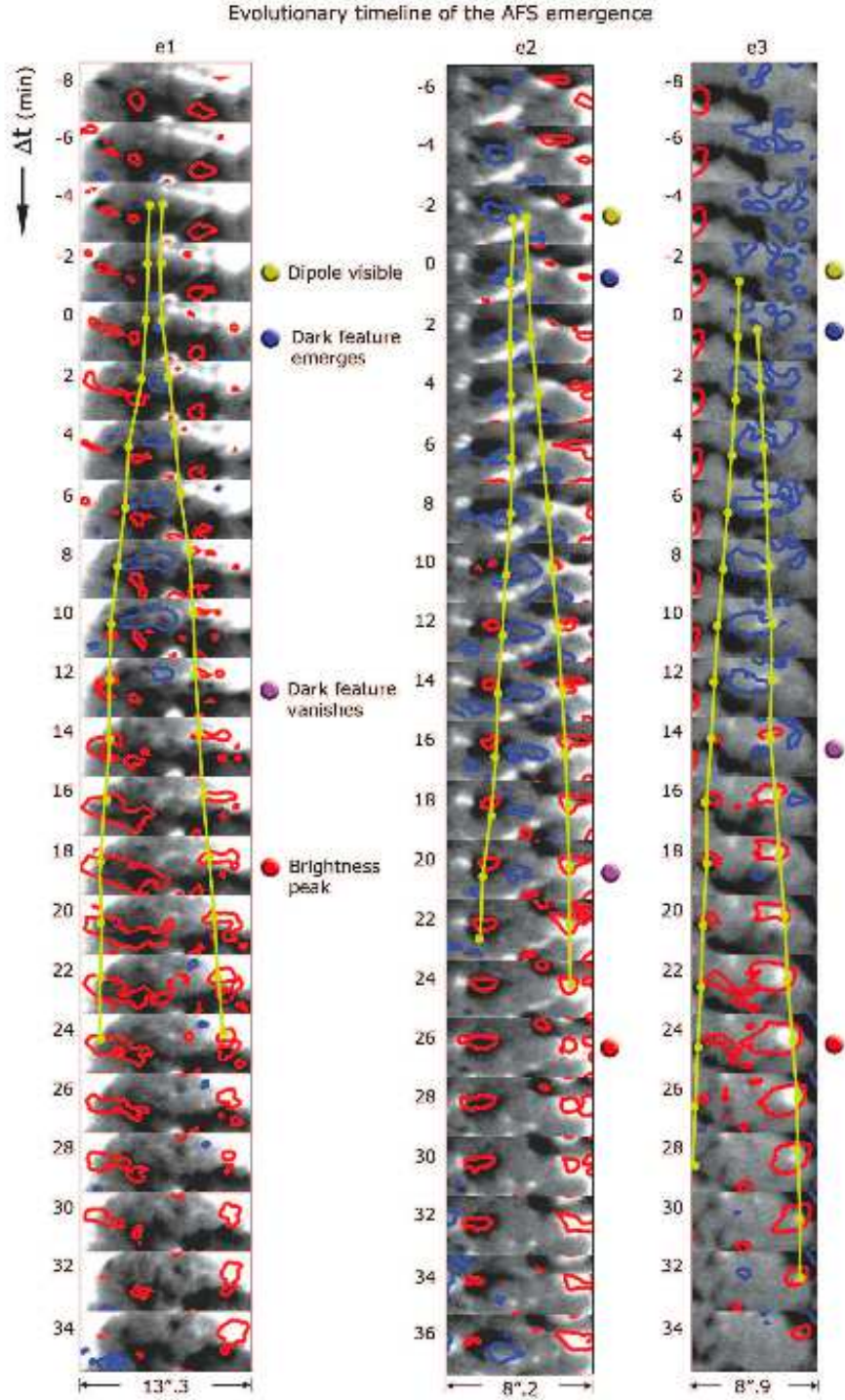} 
\caption{Timeline describing the observed evolution of the AFS emergence events through sequences of contrasted Stokes $V$ magnetograms (columns e1, e2, and e3).  The FOV  in all three cases correspond to the framed regions in the last image of Stokes $V$ sequences in Figures~\ref{secuencia1}, \ref{secuencia2} and \ref{secuencia3}. Blue/red contours correspond to dark/bright structures in the simultaneous Ca {\sc ii}~H images. The emerging bipole (both polarities) is denoted by yellow dots and connected by yellow lines. Milestones (coloured bullets) are located at specific times and refer to an observed process. It is noteworthy that all three elementary bipoles follow Hale's law, i.e. the leading polarity is positive, the following one is negative, just as in the parent AR. Movies are available in the electronic version.}
\label{timeline}
\end{figure*}

Due to the very small scale of the features we are searching for, we extract regions within the FOV in Figures~\ref{secuencia1}, \ref{secuencia2} and \ref{secuencia3}  (rectangular boxes in the last image in the  corresponding NFI V/I sequence). Moreover we enhance the circular polarization signal by clipping the images. Figure~\ref{timeline} shows the sequences for events e1, e2, and e3 (in columns) with 22 frames each. The timeline is set as in Figures~\ref{secuencia1}, \ref{secuencia2} and \ref{secuencia3} in which  $\Delta t$=0 is the reference taken at the moment where the dark features are first observed. Note that for e2 the time label for the initial frame ($\Delta t$=-6 min) is different from that in the other events though the total duration in all cases is 42 min. Red and black contours outline bright and dark areas in the simultaneous Ca {\sc ii}~H images.

The extracted regions are characterized by small-scale features and already from the first image in each episode small brightenings and darkenings in the chromospheric Ca {\sc ii}~H line are detected. Signs of a bipolar feature with positive and negative circular polarization cores start to be visible between 2 and 4 min before the appearance of the dark features described in Section~\ref{S:description}, at about the same time when dark photospheric lanes appear in Stokes $I$. A bipole associated to the SAFS is identified by yellow dots in every case (left dot being the negative circular polarization feature and right dot the positive one in all events)\footnote{Running difference images of longitudinal magnetograms generated every 2 min helped us to identify the magnetic footpoints, specially in those places where the emerging magnetic concentrations overlap the pre-existing fields and so their visualization becomes more difficult.}.  

Color buttons and text on the right of sequence e1 highlight important stages of the SAFS emergence. The first appearance of the bipole is detected about 2 min before the dark features start to be visible.  Only in one case (e3) the negative footpoint is observed prior to its positive counterpart. 
However, the signals are very weak at this initial stage and it is likely that they are below the noise level of our observation thus their limited detection. 

\begin{figure*}
\centering
\includegraphics[angle=0,width=.73\linewidth]{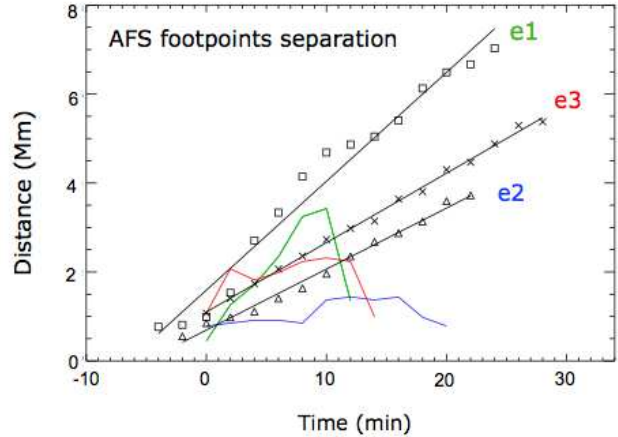} 
\caption{Separation of the footpoints of the bipolar features with time (observed on V/I circular polarization images) associated with  the emerging AFS episodes is plotted for all the three studied events: e1 (squares),  e2 (triangles),   and e3 (crosses). Black lines are the result of linear fits applied independently to the obtained values for each event. The length of the dark features in Ca {\sc ii}~H is also measured and displayed in color lines for e1 (green), e2 (blue) and e3 (red).}
\label{dipoleseparation}
\end{figure*}

From the appearance of the bipole the distance between the positive and negative fotpoints increases (see the yellow lines connecting the yellow dots in each individual sequence). In Figure~\ref{dipoleseparation} we plot the footpoint separation as a function of time for e1 (square symbols), e2 (triangle symbols) and e3 (cross symbols). Black lines result from a linear fitting procedure applied 
to the values corresponding to each of the three events. The distance between the footpoints increases constantly at rates of 4.0, 2.3 and 2.6 km s$^{-1}$  for e1, e2 and e3 respectively. The corresponding mean velocity of each footpoint then ranges from 2.0 to 1.3  km s$^{-1}$ in our sample. These values are a bit smaller than the one found by \inlinecite{mmartinez2009} of 2.95 km s$^{-1}$ yet these authors study the emergence of loops in quiet sun regions. We also compute the maximum length of the dark features outlined by the blue contours in Figure~\ref{timeline} and overplot them in Figure~\ref{dipoleseparation} for e1 (green), e2 (blue) and e3(red). The colour curves are in all cases below the corresponding lines for the separation of the footpoints (except for the red one at  the first stage of evolution) so that the footpoint separation represents an upper limit for the size of the expanding dark feature, as expected when the top of a rising loop reaches the chromosphere whereas the footpoints are further apart in the photosphere.

\begin{figure*}
\centering
\includegraphics[angle=0,width=.73\linewidth]{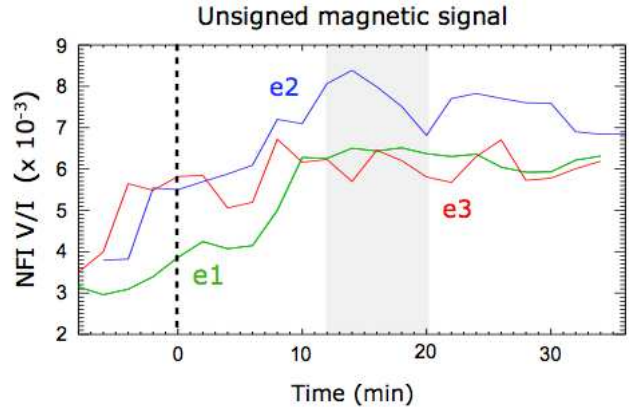} 
\caption{Magnetic evolution of the emerging bipoles associated with the small-scale AFS  for events e1, e2 and e3 shown in Figure~\ref{timeline}. The plot shows unsigned magnetic signals as computed from the NFI Stokes $V$/$I$ images. Time in the horizontal axis is the same as in Figure~\ref{timeline} where zero (vertical dashed line) corresponds to the onset of the observed dark chromospheric absorbing feature.  The grey region from 12 to 20 min extracts the time in which the dark absorbing features vanish for all the three events.}
\label{dipoloevol}
\end{figure*}

Figure~\ref{dipoloevol} shows the evolution of circular polarization signals in the footpoints of the emerging SAFS.  Unsigned circular polarization 
values are shown for all the events.  In the plot we detect the appearance of the emerging bipole followed by an increase of the unsigned signal with time. Unsigned flux of the three bipoles are similar within a 25$\%$ margin.
According to our observations the disappearance of the bipole starts to occur when one of the footpoints:  1) merge with a same-polarity feature ({\it e.g.} both polarities of e1 and the negative polarity of e3), 2) cancel  with an opposite-polarity feature ({\it e.g.} the positive polarity of e3), as visually detected over the time series. Such features correspond to more stable elements located in the surrounding area. 

It has to be pointed out that magnetic field measurements are done at the limit of detectability, so errors must be present breaking the theoretical balance in the opposite polarities of an emerging bipole. Furthermore, for the computation of the total unsigned flux, when a bipole appears in a pre-existing positive or negative environment, or when corresponding polarities from the bipole end up merging with ambient fields, cancellation may break the balance and it becomes more challenging to isolate the like polarity.

All three cases in Figure~\ref{dipoloevol}  show similar trends in the evolution of the unsigned flux and agree with a diverging bipolar structure.

\section{Discussion and Conclusions}
\label{S:discussion}

In this work we discovered small-scale dark patches emerging in the chromosphere in the data set from {\it Hinode} on  4 July 2009. We identify and follow the complete evolution of three 
events that are located in the middle portion between the 
two main emerging opposite magnetic polarities of the active region, though one is located in the dominantly positive (e1), another in the dominantly negative polarity (e3), while the third is in the vicinity of the PIL (e2). The dark features have 
remarkably similar properties as described in Sections~\ref{S:description} and \ref{S:properties}. They appear and 
evolve in regions with close-to-zero signal in the longitudinal magnetograms, i.e. weak vertical magnetic fields. These regions are typically populated with 
nearly horizontal (azimuthal) fields in which small-scale serpentine fields consisting of small-scale emerging segments ($\Omega$-loops) and U-loops \cite{spruit1987} resulting from the interaction of convective downflows and the rising emerging flux tube. The photospheric footpoints of these serpentine magnetic lines correspond to small-scale bipolar features that are observed in longitudinal magnetograms. In some cases observations show the positive and negative poles 
getting closer to an opposite polarity concentration, which could be explained as resulting from U-loop emergence {\cite{pariat2004, kubo2010}.

Further analysis of photospheric data on 4 July 2009 in the regions where we found the dark features permitted the identification of circular polarization signals corresponding to an emerging bipole. The identification of the footpoints of the emerging bipole is more difficult at the early stages of emergence due to the weak signals. The separation of the footpoints increases at the photospheric level accompanied by activity detected at the lower chromosphere. The length of the SAFS of a few Mm corresponds to the diameter of a couple of granules,  therefore they represent  granular-scale flux emergence episodes. The scale of these elementary bipoles agree well with the typical scale found by \inlinecite{strous1999} and \inlinecite{pariat2004} for serpentine flux emergence. 
It is noteworthy that the magnetic orientation of all three elementary bipoles are the same as that of the parent active region: leading polarities are positive, following ones are negative. This provides further support of a serpentine flux emergence scenario.

By carefully looking at the Stokes $I$ time series one can also identify the appearance of dark features that show up very clearly. They seem to appear 2-4 min prior to the Ca II dark features. The Stokes $I$ dark feature is the signature of the flux tube breaking through the photosphere, corroborated by theoretical simulations and observations. This provides further support to the flux emergence scenario that we propose below.

Visual inspection of the events evidences localized chromospheric Ca {\sc ii}~H brightenings appearing around the footpoints. Brightenings then expand in the direction
connecting the two footpoints, reach peak intensity, and cease. Strong intensification of brightness starts soon after the dark features vanish, indicating that chromospheric emission is 
deeply linked to the disappearance of the dark features. Noticeably, the intensity reaches a maximum value about 6 min (for e1 and e2) and 10 min (e3) after the dark features vanish. At the end of the observations, about 34 min after the starting point, the measured bright and dark intensities tend to 
return to the initial values though the regions are continuously affected by new small scale processes consequence of the highly dynamic activity. Figure~\ref{timeline} shows the increase of chromospheric activity (brightness; red contours) boosted by the bipole emergence. After a time interval of the order of 15 min magnetic fragments are observed 
to merge with major magnetic clusters and hence being incorporated in more stable magnetic structures. We evidence reconfiguration of small-scale magnetic fragments involving: (1) intensification of equal-polarity elements or (2) cancellation of opposite-polarity elements ({\it e.g.} in the positive polarity of e3). 

The level of darkness of the detected chromospheric features are not considerably greater than what is found in the interior of granular cells. The main
difference is in the coherence of the dark features, sizes and lifetimes. While dark structures in granular cells evolve fast (from seconds to a few minutes), the features we detected last for more than 10 min. But the main and more remarkable difference is the associated brightenings detected (only) at the location of the dark features we explore in this work, boosted by the interaction of emerging and ambient fields. This is the main subject we want to stress in this work and the first sign we detected in the time series that allowed the ulterior identification of the dark chromospheric features.

We also detect converging or adjacent bipoles in these sequences.  In our case e2 such adjacent opposite polarities appear at both ends of the emerging loop. Brightenings in between the opposite polarities are reminiscent of Ellerman Bombs (EBs) in shape and location. \inlinecite{pariat2004} has shown that EBs are co-spatial with bald patches ({\it i.e.} where field lines are tangent to the photosphere) in a serpentine magnetic field and suggested that EBs are due to magnetic reconnection low in the chromosphere, which leads to energy release and  also helps the underlying U-loops to ``cut off'' their heavy material-loaded sub-photospheric part and let it sink.  The fact that these U-loop bipoles all have their polarities in reverse direction to that of the emerging events provides further evidence of the serpentine flux emergence scenario.

Events like the ones described in this work are found in the region between the two main magnetic polarities, in the vicinity of the polarity inversion line, where the newer flux is in emergence. Events are not ubiquitously distributed in the active region though. Emerging SAFSs are part of serpentine-type fields that are re-arranged by reconnection to form the large-scale field of the active region. Brightenings produced by the interaction/reconnection of fields are hereby preferably located in these regions where emerging loops are more likely to interact with ambient fields, and thus mainly detected in the region between the two main magnetic polarities.

It is important to point out that these are not easy-to-detect events and that we had to play back and forth the sequence many times besides varying the contrast to be able to identify the very faint appearance of the bipole in the Stokes $V$ maps. There must be more of these events as expected from the undulatory behaviour of the emergence in this active region \cite{valori2012}. They should populate the region 
of the two magnetic tongues (formed by the azimuthal field component of a globally twisted flux tube rising through the photosphere). The three particular events we select and analyze are very distinctive and complete examples with a strong activation of chromospheric energy release (brightenings) that enhance the granular-scale AFS (SAFSs) associated with  the bipoles.

From the analysis of the observed episodes and building on previous knowledge on serpentine flux emergence, e.g \inlinecite{strous1996}, \inlinecite{strous1999}, \inlinecite{pariat2004}, \inlinecite{cheung2008}, \inlinecite{fischer2009}, \inlinecite{pariat2009}, \inlinecite{valori2012}, we can infer an evolutionary picture of the detected phenomena. 
The scenario corresponds to the following steps, which are illustrated by a sequence of cartoons in Figure~\ref{cartoons}.

\noindent 1) Horizontal magnetic flux tubes are 
below the visible photospheric level. Large loops connecting opposite magnetic polarities are observed in
the FOV and others extend beyond and reach higher atmospheric layers. Multiple scales of magnetic field concentrations are associated with the emerging flux region and both magnetic polarities can be found in all parts of the active region with an intricate fine structure which is also highly dynamic. This is a common phenomena for an emerging active region that shows a high degree of fragmentation, as pointed out by \inlinecite{strous1996}, \inlinecite{strous1999}, \inlinecite{magara2008}.\\
2) Through an interplay between buoyancy and turbulent convective motions, which deform the flux tubes and push some portions upwards, the top part (i.e. the apex) of small magnetic loops start to emerge from beneath the visible surface as manifested by the appearance of dark photospheric lanes. Some authors, {\it e.g.} \inlinecite{cheung2008}, have modelled the way rising flux tubes undulate to form serpentine field lines that emerge into the photosphere. This also agrees with previous detection of darkenings interpreted as crests of undulatory field lines penetrating the photosphere \cite{strous1999}.\\
3) Dark photospheric lanes  become longer and we detect the corresponding bipole's footpoints in circular polarization. Magnetic field presumably strengthens through convective collapse, {\it e.g.} \inlinecite{fischer2009}. Previous to the first appearance of circular polarization signals, there must be a patch of linear polarization above a granular cell, as detected by \inlinecite{mmartinez2009}, that unfortunately we can not discern due to the lack of corresponding data.\\
4) Cool dense material contained in the rising loops reach the chromospheric level where they are seen as an expanding dark feature a couple of minutes after step 3). The circular polarization signals, that correspond to the legs of the emerging loop, are observed to separate with time. This is in agreement with the detection of emerging field lines moving upward to higher atmospheric layers \cite{mmartinez2009}.\\
5) Chromospheric loops get longer and thicker as the underlying bipole's opposite polarities diverge. The dense material is proposed to be drained along the loops.\\
6) Some heating starts due possibly to reconnection with overlying pre-existing fields (i.e. component reconnection due to the angle of the serpentine fields and the overlying field) and the cool material brought up from the (sub)photospheric layers will start to be heated up, so the absorption dark feature (SAFS) disappears at about 14 min after step 3). Reconnection process has been exhaustively studied and shown to play a key role in the evolution of active regions at different scales \cite{pariat2009,harra2010,valori2012}.\\
7) Particles are accelerated in the reconnection process that takes place with ambient field.  These particles spiral downwards along the loops, impact denser layers,  decelerate and eventually heat those layers by transferring their kinetic energy.\\
8) Some evaporation of the chromosphere is initiated by step 7) and hence loops start to be filled up with heated bright plasma. Bright regions extend upwards from the footpoints 
along the emerging loops. The increase of chromospheric emission is very rapid and it lasts for about 5 min. 
The figure represents also the resistive emergence scenario \cite{pariat2004,pariat2009} triggered by magnetic reconnection just above the dipped lower part of the represented field lines.\\
9) Loops end up reconnecting with opposite-polarity magnetic field or become part of stronger pre-existing same-polarity entities \cite{valori2012}. It takes something like 35 min for the loops to complete all the steps: Emerge from below the visible photosphere and become visible in circular polarization signals. The finding that the magnetic imbalance in these elementary bipoles seem to follow that of their magnetic environment may suggest pre-emergence of fast-acting real-time cancellation processes. Resistive emergence of undulatory flux tubes allows the leftmost part of the flux to emerge (evidence of Ellerman bombs in lower atmospheric layers), and submerges the remaining lower portion. Succesive reconnections permits the dense plasma to sink and allows the rise of magnetic flux through upper atmospheric layers as pointed out by \inlinecite{pariat2004}.\\ 


\begin{figure*}
\centering
\includegraphics[angle=0,width=0.95\linewidth]{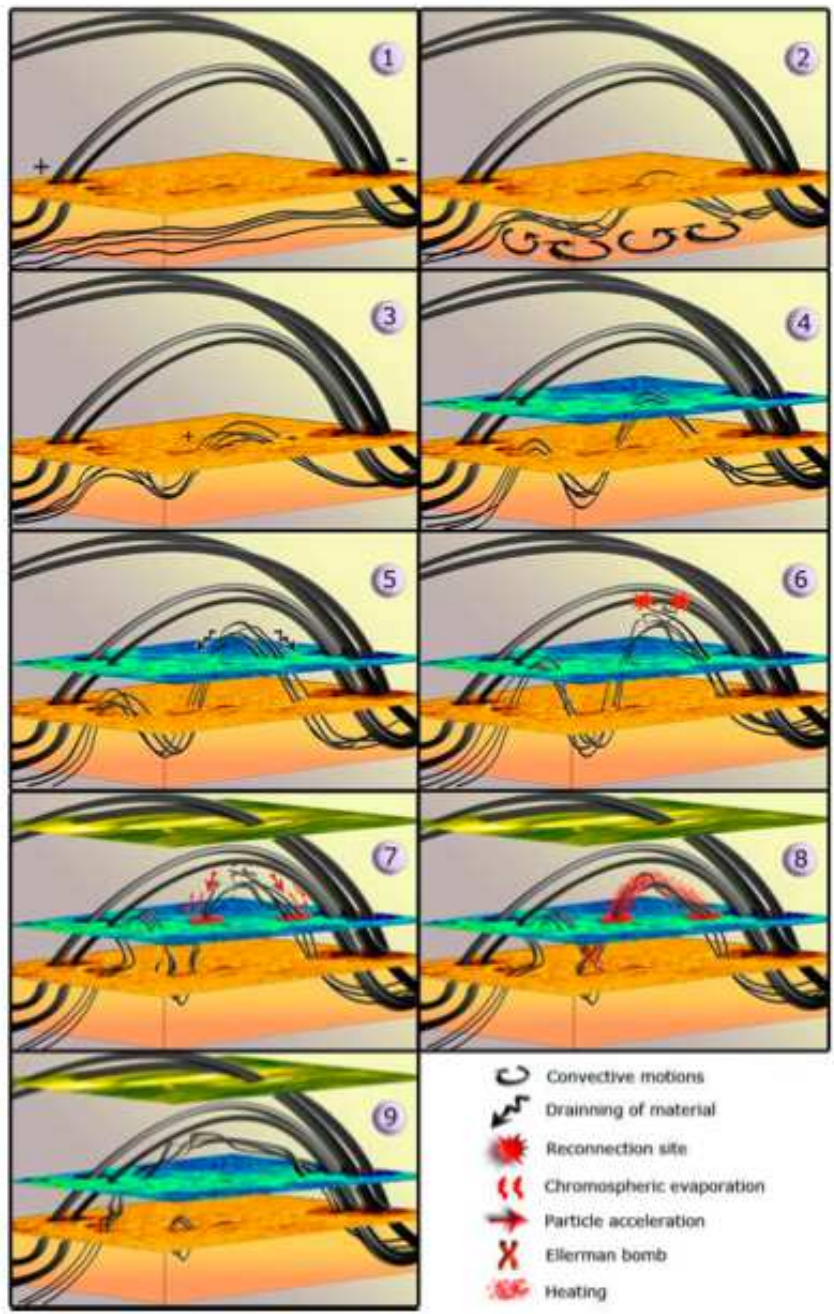} 
\caption{Sequence of cartoons showing the evolution of a flux emergence episode (not scaled). See the text for the explanation of each step. Movie available in the electronic version.}
\label{cartoons}
\end{figure*}

Photospheric Doppler velocities in Figures~\ref{secuencia1}, \ref{secuencia2} and \ref{secuencia3} display a large number of small upflows and downflows that revealed activity mainly influenced by convective photospheric turbulent motions. Localized upflows are noticeable at the bipoles' footpoints, as the emerging loops dragged material upwards. This material is also affected by gravitational forces and could be partially drained back to the surface.
The situation is yet far from being simple as newly emerging loops and dragged material are likely to encounter the already falling material in neighbouring loops which overlap along the LOS  (i.e., the total LOS integration of Doppler signals can vary significantly and even cancel out as given by upward and downward contributions from independent flux tubes). Furthermore, p-mode oscillations are also present, making the maps too noisy to show any consistent patterns.
 
The analysis of the emergence episodes give us some clues on the heating process taking place in these regions. Heating is likely to occur 
triggered by reconnection. Reconnection is expected as emerging loops rise into a pre-existing magnetic field. Different types of magnetic reconnection take place associated to the small-scale flux emergence: a) component reconnection that may involve merging of same magnetic polarities, b) U-loop emergence triggering bald-patch reconnection and subsequent energy release from cancellation processes (Ellerman Bombs)  and  c) genuine cancellation of opposite magnetic polarities not belonging to the same serpentine flux tube.  

Effects of the emergence of the SAFS do not seem to protrude into the corona though we are limited by the spatial and temporal resolution of the instruments (i.e \emph{Hinode}/EIS and XRT) to detect such small-scale phenomena. We also lack data from upper chromospheric lines (i.e H$\alpha$) that would give us a picture of the response, if any, beyond the lower chromospheric layers. According to our observations, multiple emergences can appear close to each other (with $\sim$3$\arcsec$ of separation) with apparent no interaction effects as far as individual dark features and associated brightenings are observed in the mean FOV of 10$\arcsec$ $\times$ 10$\arcsec$  as independent  events.

The same observational picture rises from the analysis of emerging small-scale loops in previous works (with comparable photospheric and chromospheric darkenings being a substantial difference the way the emerging fields interact, i.e. when in active regions emerging loops experience reconnection with ambient fields while in quiet sun regions they do not.} Very small-scale flux emergence episodes are, for instance, detected in the quiet sun internetwork.  In the {\it Hinode} data set  analyzed by \inlinecite{mmartinez2009} dark features are seen in the chromospheric Ca {\sc ii}~H with similar properties as the ones we report in this work yet with a shorter lifetime of the order of 5 min and length scale of 1 to 2 Mm. These authors make use of linear polarization magnetic data to detect the very initial stage of emergence when the apex of the loops first reach the visible surface, even before the circular polarization signals from the footpoint (vertical fields) are detected. Nonetheless, in quiet sun observations the results point towards the emergence of individual loops rather than of more complex SAFS as reported in this work. Brightenings are also confined to the location of the footpoints 
and are not observed extending towards the other magnetic footpoint  leading to brightness  enhancements at the apex of the loops.  The directivity of small-scale flux emergences found in the AR studied in this work is markedly different from internetwork emerging bipoles, which show no preferential orientation \cite{dewijn2008}. Some of the loops discovered in quiet sun regions in the above-mentioned works are observed reaching upper chromospheric layers.  This suggest that reconnection at lower chromospheric layers might be less effective due to the loops entering a region with less ambient field that what is normally found within active regions.  Physical process of granular-scale emergence of magnetic loops seems to be essentially the same in quiet sun and active regions and reduced chromospheric emission observed in quiet sun is attributed to reconnection less likely to occur. Further investigations should be carried out to conclude on this fact from the analysis of quiet sun and solar active regions observations and to determine more potential differences in the emerging process.

Observations at a higher cadence would fill the temporal gaps enabling us to better detail the emergence of these SAFS. Including data from upper chromospheric layers would be important to determine the actual height coverage of these emergences and whether they can influence pre-existing magnetic fields present in those top layers.  In reconnection sites jets could be triggered and eventually reach the corona and so likely to be detected with {\it Hinode}/EIS and XRT instruments. For an example of plasma flows likely resulting from reconnection between serpentine loops and ambient field, (see; \opencite{valori2012}). Spectro-polarimetric analysis of the solar chromosphere would contribute revealing the magnetic changes resulting from these highly dynamic events and their importance in the global flux emergence and energy budget in solar active regions.

\begin{acks}
We thank the referee for valuable comments that helped to improve the presentation of our results. Special credit goes to Pascal D\'emoulin for contributing significantly to enrich the contents of the paper. This work benefited from discussions with Etienne Pariat and Arkadius Berlicki. SVD thanks Judith Palacios for fruitful discussions. SVD  acknowledges support from STFC. LvDG acknowledges funding through the Hungarian Science Foundation grant OTKA K81421, and the European Community's FP7/2007 2013 programme through the SOTERIA Network (EU FP7 Space Science Project No. 218816). This work has been supported by the Spanish Ministerio de Ciencia e Innovaci\'on through projects PCI2006-A7-0624 and  AYA2009-14105-C06-06, and by Junta de Andaluc\'{\i}a through project P07-TEP-2687.  \emph{Hinode} is a Japanese mission developed and launched by ISAS/JAXA, collaborating with NAOJ as a domestic partner, NASA and STFC (UK) as international partners. Scientific operation of 
the \emph{Hinode} mission is conducted by the \emph{Hinode} science team organised at ISAS/JAXA. This team mainly 
consists of scientists from institutes in the partner countries. Support for the post-launch operation is provided 
by JAXA and NAOJ (Japan), STFC (UK), NASA (USA), ESA, and NSC (Norway). 
 \end{acks}

\end{article} 
\end{document}